\documentclass{acm_proc_article-sp}

\begin{document}

\title{A Cooperation Model Towards the Federated Internet of Applications}

\numberofauthors{1} %  in this sample file, there are a *total*
\author{
% 1st. author
\alignauthor
Zhe Zhang, Xudong Liu, Hailong Sun, Richong Zhang, Fei Wang\\
       \affaddr{Institute of Advanced Computing Technology}\\
       \affaddr{Beihang University}\\
	\affaddr{Beijing, China}\\
       \email{zhangzhe,liuxd,sunhl,zhangrc,wangfei@act.buaa.edu.cn}
% 2nd. author
%\alignauthor
%G.K.M. Tobin\\
%       \affaddr{Institute for Clarity in Documentation}\\
 %      \affaddr{Dublin, Ohio 43017-6221}\\
%       \email{webmaster@marysville-ohio.com}
% 3rd. author
 % use '\and' if you need 'another row' of author names
% 4th. author
%\alignauthor
%G.K.M. Tobin\\
%       \affaddr{Institute for Clarity in Documentation}\\
%       \affaddr{Dublin, Ohio 43017-6221}\\
%       \email{webmaster@marysville-ohio.com}
%\alignauthor
%G.K.M. Tobin\\
%       \affaddr{Institute for Clarity in Documentation}\\
%       \affaddr{Dublin, Ohio 43017-6221}\\
%       \email{webmaster@marysville-ohio.com}
% 6th. author
}

% Just remember to make sure that the TOTAL number of authors
% is the number that will appear on the first page PLUS the
% number that will appear in the \additionalauthors section.

\maketitle
\begin{abstract}
As Internet is changing from network of data into network of functionalities, a federated Internet of applications, that every application can cooperate with each other smoothly, is a natural trending topic. However, existing integration techniques did not pay enough attention to multiple control domains for participants, i.e. application providers and end-users. In this study, we advocate a global cooperation model for all the participants counts. In particular, we propose a hybrid model to manage the cooperation among applications to achieve more optimized allocation of efforts, which means users perform lighter actions and application providers concerning less uncontrollable information. In addition, we implement the required system and show a case study which demonstrates the effectiveness of this model.
\end{abstract}

\keywords{Internet of Applications,  Multiple Participants, Cooperation Model} % NOT required for Proceedings

\section{Introduction}

%Known
With the development of Internet, an increasing number of providers deliver services through their web or mobile applications. These applications are sufficing diverse demands of users. Since users usually access some naturally related applications for tasks, automatically information transferring among applications will greatly facilitate users. Moreover, for applications providers, serving all the tasks by themselves is impossible. Thus, we need a federated Internet of applications that the applications can easily integrated.
%Unkown
%However,while people enjoying the seamless integrated applications provided by [Googles],
%Howerve, the intergration among the applications from different providers is still a challenge.
%解释问题严重性:一个公司的产品可以很好的相互配合，而不同公司的不可以，因为一个公司有统筹全局的设计者，每个产品的设计都可以为协同服务，而不同公司很难做到。
Internet scale service integration is a rising topic derived from service computing, works like Internet Service Bus (ISB), Complex Event Processing(CEP) and Message Oriented Middleware (MOM) built solid communication techniques over Internet\cite{ref:cloud}. There are also some End-user Development products like IFTTT\footnote{IFTTT, http://www.ifttt.com. A service allowing users to create connections with "if this than that" statements} in the industry.
%问题是：怎样的机制可以让不同公司的产品配合起来形成联合体。

%有的关注单个厂商开发，有的关注用户开发，有的只做入口。怎样的机制可以让不同公司的产品配合起来形成联合体？ 他们做了一些努力，却不够好。仍然处于初级阶段，仅仅技术上可链接，却实际成功很麻烦。
These solutions merely focus on improving the development for either application providers or end-users. However, a cooperation over Internet involves an end-user and many application providers, a single participant cannot undertake this responsibility comprehensively. In other words, it costs intolerable efforts for some participants to achieve the tasks. So these solutions either cannot meet user's long tail cooperation requirements or cannot have high quantity services for the public. %But we observed that Internet of Applications is a huge system with lots of application providers and end-users involved, %Since配合得更好意味着对用户来说更多的事可以自动化地完成，对应用来说他们之间可以方便地消息交换， %task是指一个用户为了完成某件事前前后后对应用功能的所有使用。
%To cooperate better means more tasks related to multiple applications can be automatically done for user, and messages can be easily exchanged between proper applications for application developers.%我们认为考量协作时每个参与者所付出的代价是非常重要的一个侧面。
%We argue that when evaluate how [smooth] such system coordinated,
%We argue that to federate the Internet of applications, it important to consider the effort of end-user and application providers, such as the complexity of actions and developing concerning objects .
%协作时付出的代价有建立时的和维护时的。
In this study, we propose a multiple participants (application providers and end-users) cooperation model.
%我们的方案是提出一个多方协作模型。这个模型下，多了一个中介者的角色，每个参与者effort可以根据knowledge灵活合理分配
In this model, responsibilities are allocated to participants, and an intermediary coordinates their works. Therefore, the efforts (such as the complexity of actions and developing concerning objects) of each participant are acceptable in more cases for cooperation requirements, such as connection determination and communication contract.%这两个问题应有专有名词.
%lengthy
%可操作地如何解决的
%We proposed a generic cooperation model for the internet of applications, along with an implementation. 
In particular, application providers submit their interfaces and requirements to the intermediary, and end-users determine the final cooperation connections with the help of the intermediary. Multiple participants are involved in this configuration process. Then, by leveraging a uniformed identification mechanism and communication protocol, applications query the intermediary and establish connection matching each end-user's preference automatically. In addition, we conduct a case study to compare the efforts in the developing process of various solutions.
%[TO INCLUDE MORE CONTRIBUTIONS AND ADVANTAGES]

%Unknown,gaps

\section{Model and Implementation}

\begin{figure}[htb]
\centering
\includegraphics[scale=0.19]{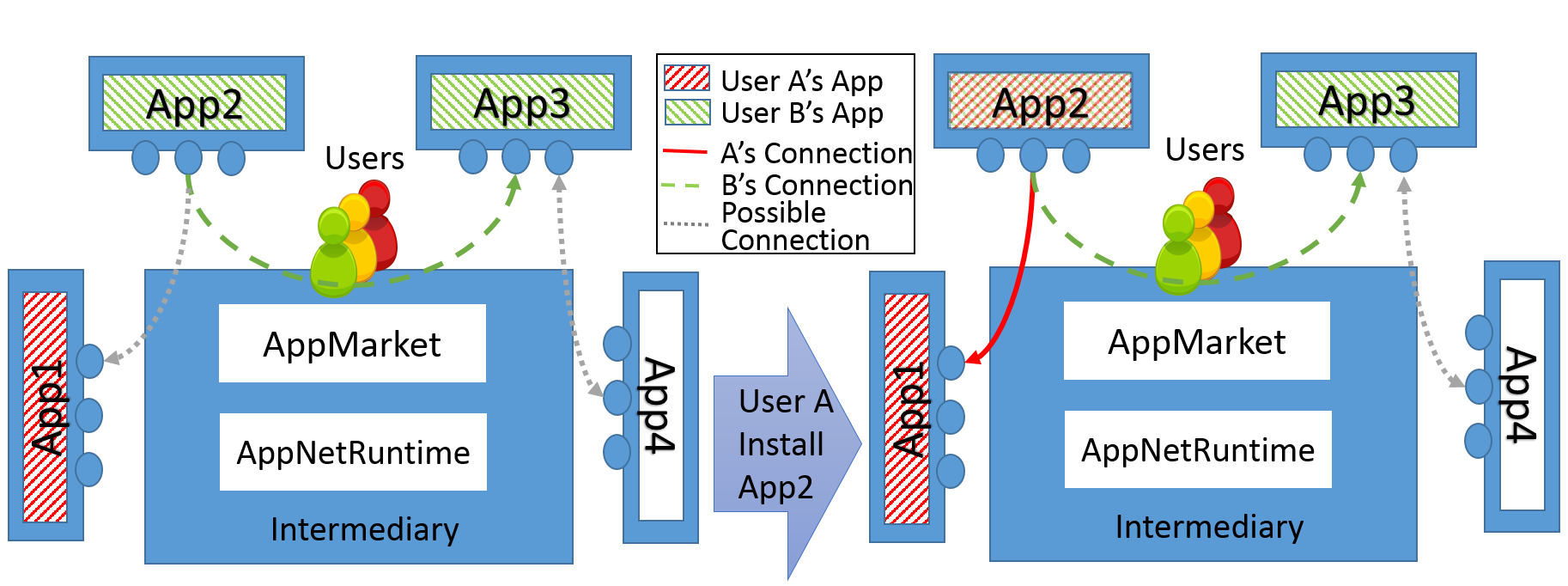}
\caption{Cooperation Model}
\label{fig:model}
\end{figure}
There are three kinds of roles in our proposed model: application, user and intermediary.
Application is an autonomous software entity raising or handling requests or events to the other entities. Application providers provide the preferences that they directly concern, typically the interface and some partial intents for the other end in a connection. Entities are denoted by boxes in Figure~\ref{fig:model}.
User means the end-users who finally be served by applications. Since end-users' preferences vary greatly due to personal interests or network environment, our model asks users to determine the connections between applications. As user is a role that cannot make real-time reaction, this fact makes our model adopt a solution that the connection is determined before the actual execution. As represented by backgrounds of apps and types of links in Figure~\ref{fig:model}, each user have his/her own network of applications and connections, which is called as \emph{userspace} in our model.
%--- generally and formally, the state collection of all the applications over the Internet for an end-user.
Intermediary is an entity that collects the preferences of applications and users to coordinate their cooperation. It is a centralized services set that carry four major duties. 1) Allowing application providers provide application definitions. 2) Giving recommendation on connections between applications. 3) Assisting users to confirm recommendations. 4) Supporting applications to query the confirmed connection and finish the final invocation.
%为什么要用户决定？

A cooperation happens in two stages: the configuration stage and the execution stage.
In the configuration stage, intermediary collects the preferences of all the participants. Applications publish or update their information to the intermediary, and the users choose the applications via intermediary. By confirming the recommendation of intermediary or connecting some applications manually, users finally determine the connection. This process is denoted by the big arrow in the middle of Figure~\ref{fig:model}.
In the execution stage, user or application raises requests by querying the connection from intermediary, and executes the final communication. 
%is user oriented and can be automatically executed.
Under the coordination of the intermediary, applications are aware of the userspace
%} --- the state collection of all the applications over the Internet for an end-user ---
and communicate as user's preferences. In Figure~\ref{fig:model}, applications communicate with each other under the connection of a particular user.
Moreover, this process leaves the data transformation among the applications to themselves, making the runtime system can be easier to be scaled.

%我们的实现，完成的这些所需做的功能，
Our implementation consists of the intermediary (AppNetRuntime and AppMarket), the protocols for connection decision and the userspace recognition, and some other auxiliary works. AppNetRuntime is a fundamental infrastructure and AppMarket is a portal web site for users. AppMarket serves for the configuration stage. Application providers submit their applications to the market along with their preferences. We exploit a named message and adapter mechanism to standardize the connection. In this mechanism, the request or the event they fire need to register a named message with data format to AppNetRuntime first, and the request or event handler should indicate some existing (or register new ones) messages they can handle. For the flexibility, they can use adapter to transform the data to their acceptable format. These information implied possible connections among applications, shown as the links in Figure~\ref{fig:model}. Then, when users install their application, our system will recommend them connections according to their userspaces for their confirmation.
AppNetRuntime is the service for the execution stage. It supports the communication among applications. We adopt an extended OAuth protocol to have the knowledge of userspace. It also responses the userspace related queries.  When an application queries the connection from it to create a communication between applications, an ID will be assigned to make the communication user-specified.

%\section{Challenges Towards The Model}
%What is the obstacles to obstacles  required for a system that realized such a model?

%\section{Our Implemetation}

%\section{Challenges and Solutions}
%To achieving such a model, there are lots of non-functional challenges to cope with.
%The problems are who will provide the intermediary service, and what is the 
%\subsection{Security and Authority}
%There are two reason for an company to 
%\subsection{Performance}

\section{Study Result}
In this section, we use a case study to show the effectiveness of our model.
Assume such a scenario: Bob lives in Beijing, and he bought a book at Amazon, latter on, he has to leave Beijing to Shanghai. He books a flight and hotel at booking.com. He hopes that Amazon can send the book to his hotel in Shanghai.
Take this task as an example, we compare our model on some metric with the existing solutions. As shown in Table~\ref{tab:comp}, the metrics including the task scope of the solution and effort cost in the preparation for different participants.
\begin{table}
\centering
\caption{Effectiveness Comparation}
\label{tab:comp}
\begin{tabular}{c|c|c|c}\hline
						& EUD 			& PD 		&MP \\ \hline
Task Coverage			& High							& Low						& Very High	\\ \hline
User Preparation				& High 							& Low 			 			& Moderate\\  \hline
Application Preparation	 	& Low							& High/Moderate						& Moderate \\ \hline

\end{tabular}
\end{table}

End-user Development (EUD) solutions, like IFTTT, allow end-users to develop some functionalities. In this case, Bob should develop an action that if he booked at booking.com then change his Amazon order, in case that booking.com and Amazon have such APIs. So long-tail requirements can be sufficed by EUD, which gives wide task coverage, but some components of task even would not exist if the requirement is not developed. It also take users some advanced efforts (such as developing the rules for IFTTT) to prepare the task.
Provider Development (PD) solutions, like Amazon's Simple Queue Service, require a provider to build the connection. PD requires Amazon have the knowledge that it will communicate with bookings.com like sites, which is very hard for provider or middleware platform (Application preparation is moderate in this situation) to know all the possible cooperation actual required by each end-user, resulting in low task coverage.
Our Multiple Participant (MP) model assigns the configuration work to all the participants: Amazon announces that they are interested in user location changed event in their applications definition at Appmarket, while bookings.com announces it will fire such events. When Bob install Amazon and bookings.com in the Appmarket, the system asks him to confirm the potential connection between them. By enabling an application cooperating with partial information and user explicitly determining connection, MP model reduces the intolerable cost for many cases, achieving wider task coverage.

\section{Discussions and Conclusions}

%what we have done
In this paper, we proposed a model to federate the Internet of applications. Realizing that the Internet of applications is evolving with various participants, we paid our attention to reduce the efforts of each participant by introducing a new cooperation model over the Internet, and built a system to verify its preliminary effect. Though it require a paradigm shifting while developing cooperative Internet applications, we belive such shifting will be the foundation of the future Internet of applications as the nature had changed.

\bibliographystyle{abbrv}
\bibliography{sigproc-sp}

\balancecolumns
% That's all folks!
\end{document}